\documentstyle[eqsecnum,aps]{revtex}
\newcommand {\be} {\begin{equation}}
\newcommand{\ee} {\end{equation}}
\newcommand{\bea}{\begin{eqnarray}}
\newcommand{\eea}{\end{eqnarray}}
\newcommand{\bean}{\begin{eqnarray*}}
\newcommand{\eean}{\end{eqnarray*}}
\newcommand{\noi}{\noindent}
\begin{document}

\draft
\preprint{hep-th/0001104}
\title{Anomalous U(1) Vortices and The Dilaton}

\author{Bruce A. Campbell and Kirk Kaminsky\\}

\address{Department of Physics, University of Alberta \\ 
Edmonton, Alberta, Canada T6G 2J1 \\}
\date{\today}
\maketitle

\begin{abstract}
\noindent
The role of the (dynamical) dilaton in the vortices associated with 
the spontaneous breaking of an anomalous U(1) from heterotic string 
theory is examined.  We demonstrate how the anomaly (and the 
coupling to the dilaton-axion) can appear in the Lagrangian and 
associated field equations as a controlled perturbation about the 
standard Nielsen-Olesen equations.  In such a picture, the 
additional field equation for the dilaton becomes a series of 
corrections to a constant dilaton VEV as the anomaly is turned on.  
In particular we find that even the first nontrivial correction 
to a constant dilaton {\it generically} leads to a (positive) 
logarithmic divergence of the heterotic dilaton near the vortex 
core.  Since the dilaton field governs the strength of quantum 
fluctuations in string theory, this runaway behavior implies 
that anomalous U(1) vortices in string theory are intrinsically 
quantum mechanical objects.
\end{abstract}


\section{Introduction}

\noi
Many four-dimensional compactifications of superstring theory 
\cite{gsw,p,bl} which preserve an unbroken N=1 spacetime 
supersymmetry 
also possess a U(1) gauge symmetry with apparently anomalous content 
for the massless fermions of the associated gauge charge.  The 
apparent anomalies of these U(1) gauge groups are canceled by a 
four-dimensional
remnant of the Green-Schwarz mechanism \cite{gs}, as originally 
argued 
by Dine, Seiberg, and Witten \cite{dsw,ads,dis}.\\

\noi
These authors noted that while the superpotential is not renormalized 
in either string or sigma model perturbation theory (so that 
solutions 
of the string equations at lowest order remain solutions to all 
orders and the vacuum remains perturbatively stable), vacuum 
degeneracy can still be lifted if a compactifcation contains a gauge 
group with an unbroken U(1) subgroup, by generating a 
Fayet-Iliopoulos\cite{fi} 
D-term.  By assumption such a term is not present at the tree level in 
the loop or 
sigma-model expansion, so the question arises as to whether it is 
possible to generate it radiatively in perturbation theory.  It turns 
out that it can arise only at one loop in the string-loop expansion, 
and 
then only if the U(1) is anomalous (since the term is proportional to 
the 
trace over the U(1) charges of the left-handed massless fermions 
\cite{ads}).\\

\noi
In fact many string compactifications have precisely such 
an anomalous U(1), with an explicit example being furnished by Dine, 
Seiberg and 
Witten for the SO(32) heterotic string.  They argue that the 
anomalies induced by 
such a U(1) are canceled by assigning the model-independent axion a 
nontrivial U(1) gauge variation, corresponding to the remnant of the 
underlying 
ten-dimensional Green-Schwarz anomaly cancellation mechanism. 
Supersymmetrically, the 
model-independent axion is paired with the dilaton [whose vacuum 
expectation value (VEV) sets the 
string-loop coupling constant] to form the scalar component of a 
chiral multiplet, whose modified (due to the anomaly cancellation and 
gauge 
invariance) Kahler potential now yields the Fayet-Iliopoulos term.  
The effect of this induced Fayet-Iliopoulos D-term, generically, is 
to break 
spacetime supersymmetry as a one-loop effect in the string loop 
expansion.  
However, the full D-term also includes contributions from charged 
scalars in the theory.  
In the known cases some of these scalars can acquire VEVs to cancel 
the Fayet-Iliopoulos D-term thereby restoring supersymmetry by 
spontaneously breaking the U(1) symmetry in a 
process referred to as vacuum restabilization.\\  

\noi
It has recently been argued that in heterotic $E_{8} \times E_{8}$ 
(as opposed to heterotic $SO(32)$) compactifications, the axion 
involved in the 
anomaly cancellation is a model-dependent axion originating from 
internal modes of the Kalb-Ramond form field $B_{ij}$, with 
$i,j=4,\ldots,9$. (The essence of this argument dates back to Distler 
and Greene \cite{dg}.)  
Such axionic modes appear paired with an internal Kahler form zero 
mode to form the 
scalar components of complex moduli $T_{i}$, which describe the size 
and shape of the compactification manifold.  However as Dine, 
Seiberg, and Witten had 
noted \cite{dsw}, if we assign one of the model-dependent axions a 
nontrivial gauge 
transformation to cancel the anomaly, and then proceed as in the 
model-independent 
case, we again get mass and tadpole terms that now appear at the 
string $\it tree$ level because there is no longer the dilaton 
(and hence string-loop) dependence that occurs in the model-independent 
case.  These terms are by assumption 
absent in the classical, massless limit of string theory.  The other 
way of 
saying this \cite{dg} is that the U(1) is not a symmetry of the 
world-sheet construction, and hence is not a symmetry of the 
low-energy 
effective theory describing the (classical) string vacuum.  
Furthermore, there is no Fayet-Iliopoulos term generated in this case, 
so spacetime supersymmetry is not 
spontaneously broken and the vacuum destabilized.  Thus, henceforth, 
we will work 
within the usual framework of Dine, Seiberg, and Witten \cite{dsw} 
and consider 
anomaly cancellation via the dilaton and model-independent axion, or 
$S$ multiplet.\\

\noi
On the other hand, it is well known that the breaking of a U(1) 
symmetry can give rise to topological defects known as Nielsen-Olesen 
vortices \cite{no}, which may appear in a cosmological context as cosmic 
strings \cite{sv}.  Bin\'{e}truy, Deffayet, and Peter \cite{bdp} analyzed
the vortices arising from such anomalous U(1) scenarios and concluded that 
there exist configurations of the axion such that some of these vortices 
can be local gauge strings, whereas for other choices of the axion 
configuration the vortices are global \cite{sv}.  However, in order to 
arrive at their final model, they freeze the dilaton to its (asymptotic) 
VEV while leaving the axion dynamical.  
Since the dilaton and model-independent axion form the scalar 
component of a chiral superfield, this Ansatz explicitly breaks 
supersymmetry as they acknowledge.  Since vacuum restabilization 
perturbatively restores supersymmetry in the resulting low-energy 
effective theory, an analysis of the vortex solutions of this effective 
theory should retain the fields required by the supersymmetry.  In 
this paper we present such an analysis, and examine the structure of 
the anomalous U(1) vortex including the dilaton as a dynamical field.\\

\noi
In order to treat the dilaton, axion, and anomaly in a systematic way, we 
show that the anomaly can be treated in the low-energy effective 
Lagrangian, 
and in the field equations, as a perturbation about the Abelian Higgs 
model and Nielsen-Olesen equations respectively.  The dimensionless
Green-Schwarz coefficient $\delta_{gs}$ will be considered as the 
perturbation parameter; in the simplified model of \cite{bdp}, wherein a 
single scalar accomplishes the vacuum restablization, supersymmetry 
(SUSY) restoration, and U(1) breaking, this parameter is of order 
$10^{-3}$.  Then, looking for static, axially symmetric (vortex) solutions
of the field equations using the standard Ansatz for the Higgs (scalar) 
and gauge fields, we show that the axion is only $\theta$ dependent 
(as \cite{bdp} obtain) and the dilaton is only $r$ dependent given the 
assumed time-independent, cylindrical symmetry of the fields. The axion 
field equation effectively decouples (we still obtain the asymptotically 
converging solution of \cite{bdp} for the axion, plus the others 
corresponding to global axionic strings), and we obtain 
ordinary differential equations (ODEs) for the dilaton, Higgs modulus, 
and the nontrivial component of the gauge field.\\

\noi
Corrections to a constant dilaton appear only at $O(\delta_{gs})$; 
at zeroth order we simply obtain the usual Nielsen-Olesen equations 
for the Higgs and gauge field.  Using a parametrization for the 
solutions to the Nielsen-Olesen equations 
correct at the asymptotic limits $r \rightarrow \infty$, and $r 
\rightarrow 0$, we obtain the first order correction to the dilaton.  
We find that the correction necessarily diverges logarithmically to 
positive infinity as $r \rightarrow 0$ as a direct consequence of 
the $r \rightarrow \infty$ boundary condition and the two-dimensional 
nature of the problem.  We also 
show this is not an artifact of the parametrization of the 
Nielsen-Olesen solutions, but 
is only dependent on these asymptotic regimes.  This divergence 
reflects a transition to a (heterotic) strong-coupling regime 
and hence a failure of the effective theory as a classical limit 
(since the 
large dilaton field means large quantum effects).  Finally, to check 
the 
consistency of this result outside of $\delta_{gs}$ perturbation 
theory, we 
examine exact solutions to the large-dilaton limit of the full 
dilaton field 
equation, which involves exponential dilaton self-couplings, and the 
axion 
contribution, neither of which is visible in the first order 
$\delta_{gs}$ 
perturbation theory.  We find the same singularity structure of the 
dilaton at $r=0$ 
as the $O(\delta_{gs})$ result, indicating a breakdown of the full 
classical approximation in the vortex core.\\

\section{Model Lagrangian}

\noi
Independently of the compactification scheme to four dimensions, the 
antisymmetric tensor field $B_{MN}$ yields via dualization the 
universal or model-independent axion $a$, which combines with 
the four-dimensional dilaton to form the scalar component of a 
chiral superfield denoted by S.  After dimensional reduction to four
dimensions, Weyl transformation to Einstein metric, and Poincar\'{e} 
duality, the relevant bosonic terms of the effective action are\cite{p}

\be 
S_{4D,het}  =  \int d^{4} x 
\hspace{4pt} (-G_{(4)} )^{\frac{1}{2}} 
\hspace{4pt} \left\{ \frac{1}{2 \kappa_{4}^{2}} 
\left[ R^{(4)}  - 2 \hspace{2pt} \frac{\partial_{\mu} S 
\partial^{\mu} S^{*}  
}{(S+S^{*})^{2}} \right] - \frac{1}{4 g_{4}^{2}}  
\hspace{4pt} \left[ e^{-2 \Phi_{4}} \hspace{2pt} F^{a\mu\nu} 
F^{a}_{\mu\nu}     
-a \hspace{4pt} F^{a\mu\nu} \tilde{F}^{a}_{\mu\nu} \right] \right\} + 
\ldots       
\label{4D-het-action}
\ee

\noi
where the ellipsis represents compactification-dependent terms involving 
the other T-like moduli of the orbifold or Calabi-Yau manifold,  
threshold corrections, and the scalars (matter fields) coming from 
the ten-dimensional gauge fields.  Here, $g_{4}^{2} = \kappa^{2}_{4}/ 
\alpha^{\prime}$, and $S = e^{-2 \Phi_{4} } + ia$ defines the 
four-dimensional dilaton and (model-independent) axion.  With respect to
the general supergravity action \cite{p}, the relevant features are 
the dilaton-axion Kahler potential given by $-\log(S+S^{\dagger})$, and 
the gauge kinetic function given by $f_{ab} = \frac{\delta_{ab}}{ 
g^{2}_{4} } S$.\\

\noi
Many compactifications of string theory possess gauge groups 
containing U(1) subgroups.  Sometimes the quantum numbers of the 
massless fermions associated with such a compactifaction appear to 
lie in anomalous representations, and hence the U(1) is referred to as 
anomalous.  As Dine, Seiberg, and Witten \cite{dsw} showed, the 
Green-Schwarz mechanism of the underlying string theories
(which ensures that the string theories themselves are anomaly free)
has a four-dimensional remnant which cancels the would-be anomalies 
associated with U(1).  Specifically, the axion-gauge coupling in
Eq. (\ref{4D-het-action}) implies that an anomalous U(1) variation,
$\delta {\cal L}_{eff} = - \frac{1}{2} \delta_{gs} \lambda 
F_{\mu\nu} \tilde{F}^{\mu\nu}$ (where $\delta_{gs}$ is the 
anomaly coefficient), can be canceled by assigning the axion 
a nontrivial U(1) variation: $a \rightarrow a + 
2 \delta_{gs} \lambda$.  In terms of the superfield S this reads:

\be
S \rightarrow S + 2 i \delta_{gs} \Lambda,
\label{dilaxion-var}
\ee

\noi
where $\Lambda$ is the supersymmetric generalization of the gauge 
transformation parameter $\lambda$.  Gauge invariance implies 
we must modify the dilaton-axion Kahler potential to

\be
K = -\log(S + S^{\dagger} - 4\delta_{gs} V).
\ee

\noi
with $V \rightarrow V + i (\Lambda - \Lambda^{\dagger})/2$ the 
vector superfield containing $A$.  Among other terms this induces a 
one-loop (in the string loop expansion) Fayet-Iliopoulos term \cite{fi}.  
Specializing to the anomalous U(1) sector of the theory, and including
the contributions coming from the (other) scalars charged under 
the U(1), denoting the 4D dilaton now by $\Phi_{4} \rightarrow \Psi$, 
and the scalar (chiral) superfields by ${\cal A}_{i}$ with charges 
$X_{i}$ and scalar components $\Phi_{i}$, we can write the effective 
Lagrangian of our model:

\be
{\cal L} = \int d^{4}\theta \left[ K(S,S^{\dagger}) + 
{\cal A}_{i}^{\dagger} e^{X_{i} V} {\cal A}_{i} \right]  + \int 
d^{2}\theta 
\frac{1}{4} S W^{\alpha} W_{\alpha} + h.c.   
\label{susy-eff-Lag}
\ee

\noi
with $W^{\alpha}$ the spinor (chiral) superfield associated with the 
field strength of V.  While a superpotential for the ${\cal A}_{i}$ 
could be added, since it must be independent of the dilaton 
superfield S in perturbation theory, we neglect it for simplicity 
since we are primarily interested in dilaton-axion dynamics.\\ 

\noi
Expanding this in component form and eliminating the auxillary
field of V by its algebraic equation of motion yields

\bea
{\cal L}_{bos} &=&  - \partial_{\mu} \Psi \partial^{\mu} \Psi
- \frac{e^{4 \Psi}}{4} \left( \partial^{\mu} a - 2 \delta_{gs} A^{\mu} 
\right)^{2} - (D_{\mu} \Phi_{i})^{\dagger} D^{\mu} \Phi_{i} \nonumber 
\\
 & & - \frac{1}{4} e^{-2 \Psi} F^{\mu\nu}  F_{\mu\nu} + \frac{1}{4} a 
F^{\mu\nu} \tilde{F}_{\mu\nu} 
- \frac{e^{2 \Psi}}{2} \left( e^{2\Psi} \delta_{gs} + X_{i} 
\Phi_{i}^{\dagger} \Phi_{i} \right)^{2}.        
\label{eff-Lag}
\eea

\noi
Equation (\ref{eff-Lag}), with the Planck mass restored everywhere 
(which we have implicitly suppressed by setting $\kappa_{4}=
\alpha^{\prime}=1$) and with $s$ instead of $e^{-2\Psi}$ for 
the dilaton, agrees with the Lagrangian of reference \cite{bdp}.
\\


\section{Perturbation Scheme and Field Equations}

\noi
In string theory the dilaton is the string loop expansion 
parameter, its vacuum expectation value setting the 
string coupling constant \cite{gsw}.  As is evident from 
Eq. (\ref{eff-Lag}), its four dimensional remnant in this model 
manifestly sets the U(1) {\it gauge} coupling: 
$\langle e^{\Psi} \rangle = g$.  Since our main interest is in 
the dilaton, it will be convenient for our purposes to 
consider variations of the dilaton about its vev.  Thus 
define $\psi \equiv \Psi - \langle \Psi \rangle$ so that

\be
e^{\Psi} \equiv g e^{\psi}.   
\ee

\noi 
We will henceforth refer to $\psi$ as the dilaton.  Then $\psi = 0 
\hspace{5pt} \leftrightarrow \hspace{5pt}  \langle Re(S) \rangle = 
1/g^{2}$. Inserting this 
into Eq. (\ref{eff-Lag}), restoring the Planck mass, and rescaling 
$\delta_{gs}$ and $a$ by $1/g^{2}$  we have

\bea
{\cal L}_{eff} & = & - M_{p}^{2}  \partial_{\mu} \psi \partial^{\mu} 
\psi - 
(D_{\mu} \Phi_{i})^{\dagger} D^{\mu} \Phi_{i} - \frac{e^{-2 \psi}}{4 
g^{2}} F^{\mu\nu} F_{\mu\nu} + \frac{a}{4 g^{2} M_{p}} F_{\mu\nu} 
\tilde{F}^{\mu\nu} \nonumber \\
& &  - M_{p}^{2} e^{4 \psi} \left( \frac{\partial^{\mu} a}{2 M_{p}} - 
\delta_{gs} A^{\mu} \right)^{2} - \frac{g^{2} e^{2\psi}}{2} \left( 
\delta_{gs} M_{p}^{2} 
e^{2\psi} + X_{i} \Phi_{i}^{\dagger} \Phi_{i} \right)^{2}.
\label{eff-Lag2}
\eea

\noi
This is invariant under local U(1) gauge transformations [with gauge 
parameter $\lambda(x^{\mu})$] which now read

\be
\Phi_{i} \rightarrow e^{i X_{i}\lambda } \Phi_{i} \hspace{10pt} , \hspace{10pt}  
A_{\mu} \rightarrow A_{\mu} + \partial_{\mu} \lambda \hspace{10pt} , \hspace{10pt} 
a \rightarrow a + 2 M_{p} \delta_{gs} \lambda.  
\label{g-trans}
\ee

\noi
As discussed above, the gauge variation of the axion in the 
$F \tilde{F}$ term cancels the anomalous variation of the Lagrangian 
due to the (suppressed) fermions.  In weakly coupled string theory, 
the anomaly coefficient $\delta_{gs}$ is calculated to be \cite{dsw}

\be
\delta_{gs} = \frac{1}{192 \pi^{2}} \sum_{i} X_{i},
\ee

\noi
where the sum is over the U(1) charges of the massless fermions and 
hence, by supersymmetry, over the charges of the massless bosons.  
In semi-realistic string models this sum may be large.  A particular 
example furnished by the free-fermionic construction \cite{fp} 
yields $Tr(Q_{X}) =  72/\sqrt{3}$, so that $\delta_{gs} \sim 10^{-2}$.
Assuming without loss of generality that $\delta_{gs}>0$, the presence
of a single scalar with negative charge can minimize the potential 
in Eq. (\ref{eff-Lag2}) (assuming we assign the other scalars zero VEVs), 
thereby canceling the Fayet-Iliopoulos D-term, restoring supersymmetry,
and spontaneously breaking the U(1) gauge symmetry.  Thus, as in 
\cite{bdp}, we consider a single Higgs scalar $\Phi$ with negative 
unit charge, effectively ignoring quantum fluctuations of the other 
scalars about their zero VEVs, and working in the classical limit.  
This is consistent with ignoring the fermionic constributions.\\

\noi
Then Eq. (\ref{eff-Lag2}) essentially becomes an Abelian Higgs model, 
coupled to the dilaton and axion through the anomaly, which may
be viewed as a perturbation.  To motivate this perspective, 
introduce a fictitious scaling parameter $\alpha$ so that

\be
\delta_{gs} \rightarrow \alpha \delta_{gs}. 
\ee

\noi
Then, as $\alpha \rightarrow 0$, the anomaly is turned off.  In order for 
the spontaneously broken Abelian Higgs model to remain in this limit, 
the invariance of the term $\delta_{gs} M_{p}^{2} e^{2\psi}$
in the potential, and in turn the gauge transformation of
the axion, imply respectively that $M_{p}$ and $a$ should scale as :

\be
M_{p} \rightarrow \alpha^{-1/2} M_{p} 
\hspace{15pt},\hspace{15pt} a \rightarrow \alpha^{1/2} a. 
\label{mp-a-scale}
\ee

\noi
Next we switch to dimensionless variables using the symmetry breaking 
scale defined by $\delta_{gs}^{1/2} M_{p}$\footnote{As typically
$\delta_{gs}^{1/2}  < 10^{-1}$, the tension of our vortex solutions, 
which is set 
by the scale of the spontaneous U(1) breaking, is below the Planck 
scale, justifying
our neglect of metric back reaction in our analysis of these 
solutions.}

\be
\hat{x}^{\mu}  =  g \delta_{gs}^{1/2} M_{p} x^{\mu} 
\hspace{8pt} , \hspace{8pt}
\hat{\phi}  = \frac{\phi}{\delta_{gs}^{1/2} M_{p} } 
\hspace{8pt} , \hspace{8pt}
\hat{A}^{\mu} = \frac{ A^{\mu}}{g \delta_{gs}^{1/2} M_{p} } 
\hspace{8pt} , \hspace{8pt}
\hat{a} = \frac{a}{\delta_{gs} M_{p}}, 
\ee

\noi
where we have written $\Phi = \phi e^{i \eta}$, so $(D_{\mu} 
\Phi)^{\dagger} D^{\mu} \Phi = \partial_{\mu} \phi \hspace{2pt} 
\partial^{\mu} \phi + \phi^{2} \left( \partial_{\mu} \eta + 
A_{\mu} \right)^{2}$.  By design, these dimensionless variables 
are $\alpha$ invariants as required for a consistent 
perturbation scheme.  Effecting these transformations and dropping 
the hats, we arrive at our final Lagrangian form:

\bea
{\cal L}^{\prime}_{eff} &=& \frac{-1}{\alpha \delta_{gs}} 
\partial_{\mu} \psi 
\partial^{\mu} \psi - \partial_{\mu}  \phi \hspace{2pt}
\partial^{\mu} \phi - \phi^{2} (\partial_{\mu} \eta + A_{\mu})^{2} 
\nonumber \\
& &- \frac{e^{-2 \psi}}{4}  F^{\mu\nu} F_{\mu\nu} - 
\frac{e^{2\psi}}{2} \left( \phi^{2} - e^{2\psi} \right)^{2} \nonumber 
\\
& & + \alpha \delta_{gs} \left[ \frac{a}{4} F_{\mu\nu} 
\tilde{F}^{\mu\nu} - 
\frac{e^{4\psi}}{4} \left( \partial^{\mu} a - 2 A^{\mu} \right)^{2} 
\right],      
\label{eff-Lag-final}
\eea

\noi
where we have rescaled the overall Lagrangian by the factor 
$M_{p}^{4} g^{2}\delta_{gs}^{2}$.  In the 
limit $\alpha \hspace{2pt} \delta_{gs} \rightarrow 0$, we identically 
get the spontaneously broken Abelian Higgs model\footnote{As we will 
later 
show explicitly, in this limit, the dilaton $\psi \rightarrow \langle 
\psi 
\rangle \equiv 0$, so its gradients vanish identially.}.  Thus,
since only the combination $\alpha \delta_{gs}$ appears, setting 
$\alpha=1$ 
(or relabeling $\beta=\alpha\delta_{gs}$), the only remaining 
parameter is 
$\delta_{gs}$ (or $\beta$) which is now to be interpreted as a perturbation 
parameter\footnote{Strictly speaking, since the $a$ defined here was 
rescaled by 
$\delta_{gs}$, $\alpha$ is the perturbation parameter.}.\\

\noi
The field equations derived from Eq. (\ref{eff-Lag-final}) are

\bea
 \Box{\psi} &=& \frac{\beta}{2} \left[ e^{2\psi} (3 e^{2\psi}-\phi^{2})
 (e^{2\psi}-\phi^{2})  - \frac{e^{-2\psi}}{2} F^{\mu\nu} F_{\mu\nu} 
\right] + \frac{\beta^{2}}{2} 
e^{4\psi} 
(\partial^{\mu} a - 2 A^{\mu})^{2} \label{d-eom} \\
\Box{\phi} &=& \phi (\partial_{\mu}\eta + A_{\mu})^{2} + 
e^{2\psi}\phi (\phi^{2} - e^{2\psi}) \label{h-eom} \\
0 &=& \partial_{\mu} [ \phi^{2} (\partial^{\mu}\eta + A^{\mu}) ]  
\label{e-eom} \\
\Box{a} &=& 2 \partial_{\mu} A^{\mu} - \frac{e^{-4 \psi}}{2} 
F_{\mu\nu} 
\tilde{F}^{\mu\nu} - 4 \partial_{\mu} \psi (\partial^{\mu} a - 2 
A^{\mu}) \label{a-eom} \\
\partial_{\mu} (e^{-2\psi} F^{\mu\nu}) &=& 2 \phi^{2}(\partial^{\nu} 
\eta +A^{\nu}) + \beta \left[ \partial_{\mu} (a \tilde{F}^{\mu\nu}) 
- e^{4\psi} (\partial^{\nu} a - 2 A^{\nu}) \right].  
\label{g-eom}
\eea

\noi
First we note that despite the presence of the dynamical dilation, by 
differentiating Eq. (\ref{g-eom}) with respect to $x^{\nu}$, and then 
using Eqs.
(\ref{e-eom}), (\ref{a-eom}), and $\partial_{\mu}\tilde{F}^{\mu\nu} = 
0$, we still obtain

\be
\tilde{F}^{\mu\nu} F_{\mu\nu} = 0.
\ee

\noi
Then, after choosing the Lorentz gauge $\partial_{\mu} A^{\mu} =0$, 
the axion field equation (\ref{a-eom}) simplifies to

\be
\Box{a} = - 4\partial_{\mu} \psi ( \partial^{\mu} a - 2 A^{\mu} ).   
\label{a-eom2}
\ee


\section{Vortex ODE's}

\noi 
It is well known that the spontaneously broken Abelian Higgs model 
possesses topologically stable vortex solutions 
sometimes called Nielsen-Olesen vortices \cite{no} (see Shellard and 
Vilenkin \cite{sv} for a complete reference on the subject).  
These correspond to static, cylindrically symmetrical 
solutions of the field equations for the Higgs and gauge fields.  
Specifically, working in cylindrical coordinates $(t,r,\theta,z)$ we 
look for solutions independent of $t$ and $z$, with the standard 
vortex Ansatz \cite{no},\cite{sv} for the Higgs phase and the gauge 
field:

\bea 
\eta &=& n \theta, \nonumber \\
A^{\mu} &=& ( 0,0,A^{\theta}(r),0) \equiv ( 0,0,A(r),0),
\eea

\noi
where n is an integer characterizing the winding number of the vortex.
The Higgs field $\Phi = \phi e^{i\eta} \rightarrow  \langle \phi 
\rangle e^{i\eta}$ (as $r\rightarrow \infty$) defines a 
representation of the U(1) gauge group 
space $S^{1}$ since from Eq. (\ref{g-trans}), $\Phi \rightarrow e^{-i 
\lambda} \Phi$ under a gauge transformation.  Thus $\Phi$ defines (as 
$r\rightarrow 
\infty$) a mapping from the boundary $S^{1}$ of physical space onto 
the group space $S^{1}$, and so can topologically be classified by an 
integer n.  In the language of homotopy theory $\pi_{1}(S^{1}) = 
{\cal Z}$.  With these Ansatze, the Higgs phase field equation 
(\ref{e-eom}) can be written as 

\be
\frac{1}{r} \frac{\partial \phi}{\partial \theta} ( \frac{n}{r}  + A) = 0,
\ee

\noi
where we have used $\partial_{\mu} A^{\mu}=0$ and the fact that 
$\eta=n\theta$ 
implies $\Box\eta = 0$.  Then since in general $A(r) \neq -n/r$, we 
get

\be
\frac{\partial \phi}{\partial \theta} = 0 \hspace{10pt} \Rightarrow 
\hspace{10pt} 
\phi = \phi(r).
\ee

\noi
This is normally assumed as an Ansatz, but this shows it actually 
follows from the Higgs phase field equation.  Then Eq. (\ref{e-eom}) is 
identically satisfied with these forms of $\eta$, $A$, and $\phi$.  
At this point we 
still have $a=a(r,\theta)$, and $\psi=\psi(r,\theta)$ assuming only 
static, axial symmetry.  However, writing the Higgs modulus equation 
(\ref{h-eom}) 
as\footnote{Remember we are always working with metric signature 
$(-,+,+,+)$ so $\Box = 
-\frac{\partial^{2}}{\partial t^{2}} + \triangle$, etc.}

\bea
\Box{\phi} - \phi (\partial_{\mu}\eta + A_{\mu})^{2} &=& \frac{d^{2} 
\phi}{d r^{2}} + \frac{1}{r} \frac{d\phi}{d r} - \phi(r) \left[ \frac{n}
{r} + A(r)\right]^{2} \nonumber \\ &\equiv& f(r) \nonumber \\
&=& e^{2 \psi(r,\theta)} \phi(r) \left[ \phi^{2}(r) - 
e^{2\psi(r,\theta)}\right]          
\eea

\noi
determines $\psi$ algebraically as a function of r alone, so  $\psi = 
\psi(r)$.  Furthermore, consider the gauge field equation (\ref{g-eom}) 
for $\nu = r$, i.e. $\nu=1$.  Since 
$A^{\mu} = \delta^{2\mu} A(r)$, only $F^{12}$ and $\tilde{F}^{03}$ 
are nonzero.  Then Eq. (\ref{g-eom}) for $\nu=1$ reads

\be
\frac{1}{r} \frac{\partial}{\partial \theta}\left[ e^{-2\psi(r)} 
F^{21}(r) \right] \equiv 0 = 2 \phi^{2} ( 0+0) + \beta \left[ 0 - 
e^{4\psi} ( \frac{\partial a}{\partial r} - 0) \right] 
\hspace{20pt} \Rightarrow \hspace{20pt} \frac{\partial a}{\partial r} 
= 0,
\ee

\noi
so that $a = a(\theta)$.  Now $\psi=\psi(r)$, $a=a(\theta)$, and 
$A=A(r)$ imply in the axion field equation (\ref{a-eom2}) that

\be
\partial_{\mu} \psi (\partial^{\mu} a - 2 A^{\mu})  = 0 \hspace{20pt}
\Rightarrow \hspace{20pt}
\Box{a} = \frac{1}{r^{2}} \frac{d^{2} a}{d\theta^{2} } = 0. 
\ee

\noi
This fixes 

\be
a(\theta) = C \theta + D. 
\ee

\noi
Because $a$ appears only derivatively coupled, we may take without 
loss of generality $D=0$.  Furthermore, single valuedness in 
the physical space requires that $C$ be an integer, so that $a$ 
represents a mapping from physical space 
into the gauge group space just as $\eta$ does (see \cite{bdp}). 
The specific axion solution of Bin\'{e}truy, Deffayet and Peter 
\cite{bdp} corresponds to the choice $C = -2n$, where n is the 
winding number of the Higgs phase\footnote{$a = - 2n\theta$ in the 
original variables reads $a=2\delta_{gs}M_{p}\eta/X$}. We will 
consider the general case for the moment, leaving $C=-2 
m$ without loss of generality ($m$ integral or half-integral), 
with m not necessarily equal to n.  Effectively this allows the 
axion and the Higgs phase to have different winding numbers.\\

\noi
Combining what we have learned about the coordinate dependences of 
the fields, we can now reduce the remaining field equations 
(\ref{d-eom}), (\ref{h-eom}), and (\ref{g-eom})  to three 
ordinary differential equations:

\be
\frac{d^{2} \psi}{d r^{2}} + \frac{1}{r} \frac{d \psi}{dr} = \frac{\beta}
{2} 
\left[e^{2\psi} (3 e^{2\psi} - \phi^{2})( e^{2\psi} - \phi^{2}) - 
e^{-2\psi} \left(\frac{1}{r}\frac{d}{dr} (r A) \right)^{2}\right]  
+ 2\beta^{2} e^{4\psi} \left(\frac{m}{r} + A\right)^{2},
\label{d-ode}
\ee
\be
\frac{d^{2} \phi}{d r^{2}} + \frac{1}{r} \frac{d \phi}{dr} = \phi 
\left(\frac{n}{r} + A\right)^{2}+ e^{2\psi} \phi 
\left(\phi^{2}-e^{2\psi}\right),
\label{h-ode}
\ee
\be
\frac{d}{dr}\left[\frac{1}{r} \frac{d}{dr}(rA)\right] = 2 \frac{d\psi}
{dr} \frac{1}{r} 
\frac{d}{dr}(rA) + 2 \phi^{2} e^{2\psi} \left(\frac{n}{r}+A\right) + 
2 \beta 
e^{6\psi} \left(\frac{m}{r} + A\right).
\label{g-ode}
\ee

\noi
As in the standard Nielsen-Olesen vortices\cite{no},\cite{sv} of the 
Abelian Higgs model, we require that the Higgs modulus approach its 
vacuum expectation value asymptotically to minimize the potential term, 
and that the covariant derivative $D_{\mu}\Phi$ vanish asymptotically 
(i.e. the gauge field asymptotically becomes a pure gauge) so that the 
energy (per unit length) of the vortex remains finite.  Translated 
into our language, these conditions read:

\bea
\phi(r) &\rightarrow& 1 \hspace{10pt} , \hspace{10pt} r \rightarrow 
\infty; \nonumber\\
A(r) &\rightarrow& \frac{-n}{r} \hspace{10pt} , \hspace{10pt} r 
\rightarrow 
\infty.
\label{asymp-cond}
\eea

\noi
The Higgs `screening' by the gauge fields prevents the logarithmic 
divergence of global vortices, so that the energy integral $\int (\frac{n}
{r}+A)^{2} \phi^{2} r dr$ (remnants of the covariant derivative 
$D_{\mu} \Phi$) is asymptotically finite.  However, after fixing
the asymptotic gauge behavior with respect to the Higgs boson, the 
presence of the axion kinetic term  $\int (\frac{m}{r} + A)^{2} r dr$ 
reintroduces these logarithmic divergences in the energy integral, unless 
$m=n$ (the result of Bin\'{e}truy et al.).  Since our primary interest is now in 
the dilaton, for the remainder of our discussion we consider the $m=n$ 
case to simplify the equations slightly.  We demonstrate in the next 
section that this will in no way affect any subsequent results.\\

\noi
Before proceeding we now make a convenient change of variables for 
the gauge field.  Define $v(r)$ through

\be
A(r) = \frac{-n [1-v(r)]}{r},
\label{v-def}
\ee

\noi
so that 
\be
v(r) \rightarrow 0 \hspace{10pt} , \hspace{10pt} r \rightarrow \infty.
\label{asympt}
\ee

\noi
Equations (\ref{d-ode})-(\ref{g-ode}) now read, denoting r 
derivatives by primes

\bea
\psi^{\prime\prime} + \frac{\psi^{\prime}}{r} &=& \frac{\beta}{2} 
\left[3 e^{6\psi} - 4\phi^{2}e^{4\psi} +\phi^{4}e^{2\psi} - 
\frac{e^{-2\psi} n^{2}}{r^{2}} (v^{\prime})^{2}  \right] + 
2\beta^{2}e^{4\psi}\frac{n^{2}v^{2}}{r^{2}}, \label{d-ode2} \\
\phi^{\prime\prime}  + \frac{\phi^{\prime}}{r} &=& \frac{n^{2}}{r^{2}} 
\phi v^{2} + e^{2\psi} \phi (\phi^{2} - e^{2\psi}), \label{h-ode2} \\
v^{\prime\prime} - \frac{v^{\prime}}{r} &=& 2 \psi^{\prime} 
v^{\prime} + 2 (\phi^{2} e^{2\psi} + \beta e^{6\psi}) v. 
\label{g-ode2}    
\eea

\noi
We require the dilaton to approach its asymptotic VEV as $r 
\rightarrow \infty$, which, in our langauge, means 

\be
\psi \rightarrow 0  \hspace{10pt} , \hspace{10pt} r \rightarrow 
\infty 
\hspace{10pt} (i.e. \langle Re(S) \rangle = \frac{1}{g^{2}}).   
\label{asymp-cond2}
\ee

\noi
Now consider the boundary conditions at $r=0$.  In the standard 
Nielsen-Olesen or Abelian Higgs model \cite{sv}, the vortex 
configuration means 
that $\phi$ attains the symmetric (false vacuum) state $\phi=0$ at 
$r=0$ (which we argued was necessary for the energy integral to be 
well defined), and $A$ remains bounded (more precisely the magnetic 
field remains bounded).  Thus we have

\be
\phi(0) = 0 \hspace{10pt} , \hspace{10pt} v(0) = 1.
\ee

\noi
This leaves, finally, the boundary condition for the dilaton at $r=0$.
Of course we would like to have the dilaton (VEV) remain bounded in 
the core, but as we shall now show, this is not possible if $\beta \neq 
0$.

\section{Perturbative Expansion and Corrections to the Dilaton}

\noi
Throughout this section we will make usage of the following 
elementary 
fact of our radial equations:

\be
f^{\prime\prime} + \frac{f^{\prime}}{r} = 0 \hspace{10pt} \Rightarrow 
f(r) = C_{1} + C_{2} \hspace{2pt} \log(r).  
\label{fund-Lap}
\ee

\noi
First, note that if $\beta=0$, then the dilaton equation 
(\ref{d-ode2}) 
becomes Eq. (\ref{fund-Lap}), so that the asymptotic condition 
(\ref{asymp-cond2}) 
on the dilaton then implies:

\be
\psi_{0} (r) \equiv 0  \hspace{10pt} \forall r.
\ee

\noi
This of course corresponds to the frozen dilaton.  Then the other two 
equations, Eqs. (\ref{h-ode2}) and (\ref{g-ode2}), identically reduce to 
the Nielsen-Olesen equations 
of the Abelian Higgs model, as promised:

\bea
\phi_{0} ^{\prime\prime}  + \frac{\phi_{0} ^{\prime}}{r} &=& 
\frac{n^{2}}{r^{2}} 
\phi_{0}  \hspace{2pt} v_{0} ^{2} + \phi_{0}  \left(\phi_{0} ^{2} - 
1\right), \label{h-no} \\
v_{0}^{\prime\prime} - \frac{v_{0} ^{\prime}}{r} &=& 2 \phi_{0} ^{2} 
\hspace{2pt} v_{0}, \label{g-no} 
\eea

\noi
with $v_{0}(0)=1$, $v_{0}(\infty)=0$, $\phi_{0}(0)=0$, 
$\phi_{0}(\infty)=1$. 
We have subscripted the fields with zeros to indicate that these are 
the zeroth order terms in a perturbation expansion in $\beta$, which 
we now define formally in the obvious way:

\be
\psi(r) = \sum_{i=0}^{\infty} \beta^{i} \psi_{i}(r) \hspace{6pt} , 
\hspace{6pt}
\phi(r) = \sum_{i=0}^{\infty} \beta^{i} \phi_{i}(r) \hspace{6pt} , 
\hspace{6pt}
v(r) = \sum_{i=0}^{\infty} \beta^{i} v_{i}(r).
\ee

\noi
Substituting these into Eqs. (\ref{d-ode2})-(\ref{g-ode2}) 
yields the following $O(\beta)$ corrections:

\bea
\psi_{1}^{\prime\prime} + \frac{\psi_{1}^{\prime}}{r} &=& \frac{1}{2} 
\left[3-4\phi_{0}^{2}+\phi_{0} ^{4} - \frac{n^{2}}{r^{2}} 
(v_{0}^{\prime})^{2}\right], \label{dil-1storder} \\
\phi_{1}^{\prime\prime} + \frac{\phi_{1}^{\prime}}{r} &=& 
\frac{n^{2}}{r^{2}} \left(\phi_{1} v_{0}^{2} + 
2\phi_{0}v_{0}v_{1}\right) + 
2\psi_{1}\left(\phi_{0}^{3}-2\phi_{0}\right)+\phi_{1} 
\left(3\phi_{0}^{2}-1\right), \\
v_{1}^{\prime\prime} - \frac{v_{1}^{\prime}}{r} &=& 2 
\psi_{1}^{\prime} 
v_{0}^{\prime} + 2 v_{0} \left( 2 \phi_{0}\phi_{1} + 2 
\phi_{0}^{2}\psi_{1} + 1\right) + 2 v_{1}\phi_{0}^{2},               
\eea

\noi
where we have included the corrections to the Higgs and gauge field 
for completeness.  What really interests us is the first of these 
equations, Eq. (\ref{dil-1storder}), the first correction to the 
dilaton.  
Note that this $O(\beta)$ correction does {\it not} depend on  
having chosen the choice of Bin\'{e}truy et al. for the axion, since the 
axion does not enter at this order.  This can be seen directly from 
Eq. (\ref{eff-Lag-final}) or (\ref{d-ode2}).  More importantly, this 
dilaton
correction can be calculated from knowledge of only $\phi_{0}$ and 
$v_{0}$, i.e. the Nielsen-Olesen solution for the Higgs and the gauge 
field.\footnote{In fact, it is obvious that the dilaton at any order 
is determined only by functions of lower order.}\\

\noi
Unfortunately explicit solutions to the Nielsen-Olesen equations 
(\ref{h-no})-(\ref{g-no}) are not known.  However, all we really need 
is a parametrization of the solutions with the correct behavior at 
$r \rightarrow \infty$ and at $r \rightarrow 0$.  The conclusions we 
will draw, will depend only on the asymptotic behavior of $\phi_{0}$, 
$v_{0}$, {\it and} in particular the $r\rightarrow \infty$ boundary 
condition on $\psi$ itself.\\  

\noi
Thus, first consider the large r behavior of the Nielsen-Olesen 
equations (\ref{h-no}), (\ref{g-no}). Write $\phi_{0}$ and $v_{0}$ 
as $1-\delta\phi_{0}$ and $\delta v_{0}$ respectively, 
where $\delta$'s represent deviations with respect to asymptotic 
values.  Then the linearizations of Eqs. (\ref{h-no}),(\ref{g-no}) are

\bea
\delta\phi_{0}^{\prime\prime} + \frac{\delta\phi_{0}^{\prime}}{r} &=& 
2\delta\phi_{0} + O(\delta^{2}), \label{h-no-asymp} \\
\delta v_{0}^{\prime\prime} - \frac{\delta v_{0}^{\prime}}{r} &=& 2 
\delta v_{0} + O(\delta^{2} ). \label{g-no-asymp}    
\eea 

\noi
Note that as per Perivolaropoulos \cite{per} (or Shellard and 
Vilenkin 
\cite{sv}), since we have the case `$\beta < 4$' (in their notation), 
we 
do not need to consider the inhomogeneous term $(\delta v_{0} 
)^{2}/r^{2}$ in the $\delta \phi_{0}$ equation, which can 
dominate a linear term of $O(\delta \phi_{0})$ if $\beta>4$.  In this 
case,
the gauge field dictates the falloff of the Higgs field.  
Our `$\beta$' (not to be confused with the perturbation parameter) is 
1, so this usual (strict) linearization applies.  The solutions to 
these linearized equations, with the asymptotic boundary 
conditions, are in terms of modified Bessel functions:

\bea
\delta \phi_{0} \rightarrow  K_{0}(\sqrt{2} r) \rightarrow C_{\phi} 
\frac{e^{-\sqrt{2} r}}{\sqrt{r}} \hspace{10pt} , \hspace{10pt} r 
\rightarrow \infty,  \label{h-larger} \\
\delta v_{0} \rightarrow K_{1}(\sqrt{2} r) \rightarrow C_{v} \sqrt{r} 
e^{-\sqrt{2} r}
\hspace{10pt} , \hspace{10pt} r \rightarrow \infty, \label{g-larger}
\eea

\noi
where $C_{\phi}$, and $C_{v}$ are constants of order 1.  As 
Perivolaropolous \cite{per} notes, the factor of $1/\sqrt{r}$ is 
usually neglected in Eq. (\ref{h-larger}).  We will neglect these 
$\sqrt{r}$ terms as being negligible with respect to the exponentials 
when parametrizing a solution of the Nielsen-Olesen equations 
over the whole range, and later argue that this does not affect our 
results.\\

\noi
Now consider the small r behavior, this time taking $\phi_{0}$ as 
$\delta \phi_{0}$.  With $v_{0}(r\ll 1) \approx 1$ the leading order 
behavior of Eq. (\ref{h-no}) at small r is

\be
\delta \phi_{0}^{\prime\prime} + \frac{\delta \phi_{0}^{\prime}}{r} = 
\frac{n^{2} \delta \phi_{0}}{r^{2} }  \hspace{10pt} \Rightarrow 
\hspace{10pt} \delta \phi_{0} = A r^{n} \hspace{10pt} , 
\hspace{10pt} r \ll 1  \label{h-smallr}   
\ee

\noi
where $A>0$ (to be determined conveniently in a moment),
and where we have discarded the second singular solution.  At this 
point we specialize to the $n=\pm 1$ vortex for simplicity.  
Then the small r gauge field equation is

\be
v_{0}^{\prime\prime} - \frac{\delta v_{0}^{\prime}}{r} = 2 
(\delta \phi_{0})^{2} v_{0} = 2 A^{2} r^{2} v_{0},  
\ee

\noi
with solution 

\be
v_{0}  =  \left. e^{-A r^{2}/\sqrt{2}}  \sim 1 - \frac{A}{\sqrt{2}} 
r^{2} + O(r^{4}) \right. \hspace{10pt},\hspace{10pt} 
r \ll 1, \label{g-smallr}  
\ee

\noi
where again we have discarded the second solution (a positive 
exponential), which has the wrong behavior near $r=0$, and used 
$v_{0}(0)=1$.  Combining Eqs. (\ref{h-larger}), (\ref{g-larger}), 
(\ref{h-smallr}),
and (\ref{g-smallr}) suggests the following parametrizations of the 
solutions to the Nielsen-Olesen equations:

\bea
\phi_{0}(r) \sim \tanh(\frac{r}{\sqrt{2}}), \label{h-par} \\
v_{0}(r) \sim \textnormal{sech}^{2}(\frac{r}{\sqrt{2}}), \label{g-par} 
\eea

\noi
which corresponds to setting $A=1/\sqrt{2}$. They have the following 
asymptotic behavior:

\bea
\phi_{0}(r) \rightarrow \frac{r}{\sqrt{2}} \hspace{12pt} (r \rightarrow 
0) \hspace{12pt} &;& 
\hspace{12pt} \phi_{0}(r) \rightarrow  1 - 2 e^{-\sqrt{2}r} 
\hspace{12pt} 
(r\rightarrow \infty) \\
v_{0}(r) \rightarrow 1 - \frac{r^{2}}{2} \hspace{12pt} (r \rightarrow 
0) 
\hspace{12pt} &;& 
\hspace{12pt} v_{0}(r) \rightarrow 4 e^{-\sqrt{2} r}  \hspace{12pt} 
(r \rightarrow \infty), 
\eea

\noi
and are therefore suitable parametrizations that become `exact' in 
both r limits.\footnote{A quick numerical check reveals that the 
error, 
by construction, is concentrated near $r=1$ and is bounded above by 
about $20\%$.}  
These are of course the usual solitonic-type forms that qualitatively 
describe the behavior of the solutions to Eqs. 
(\ref{h-no}),(\ref{g-no}) very 
well, as can be checked by comparing them with the exact numerical 
calculations.\\

\noi
Inserting Eqs. (\ref{h-par}) and (\ref{g-par}) into the dilaton correction 
(\ref{dil-1storder}) yields, after some trigonometric simplifcation,

\be
\psi_{1}^{\prime\prime} + \frac{\psi_{1}^{\prime}}{r} = 
\textnormal{sech}^{2}(\frac{r}{\sqrt{2}}) + 
\textnormal{sech}^{4}(\frac{r}{\sqrt{2}})
\frac{ \left[\textnormal{sech}^{2}(\frac{r}{\sqrt{2}}) - 
\left(1-\frac{r^{2}}{2}\right) \right]}{r^{2}}  \equiv f(r).
\label{dil-nexttofinalcorr}
\ee

\noi
However, the inhomogeneous right hand side is well approximated  
globally by the first term $\textnormal{sech}^{2}(r/\sqrt{2})$.  In 
particular, 
the dominant asymptotic behavior as $r\rightarrow \infty$ is the 
same [since the latter term is a correction of $O(\exp(-2\sqrt{2}r))$ 
coming from the $(v_{0}^{\prime})^{2}$ and the $\phi_{0}^{4}$ 
contributions], and is 
correct to $O(r)$ in the small r limit.\footnote{Alternatively, we do 
not have to make 
this truncation, at the price of making the subsequent analysis 
much more algebraically tedious, without qualitatively changing the 
result.  The point is that it will be the dominant asymptotic 
behavior that determines the dilaton behavior.}  Thus we take

\be
\psi_{1}^{\prime\prime} + \frac{\psi_{1}^{\prime}}{r} \simeq 
\textnormal{sech}^{2}(\frac{r}{\sqrt{2}}) \hspace{10pt} (\rightarrow  4 
e^{-\sqrt{2}r} 
\hspace{5pt} \textnormal{as} \hspace{5pt} r\rightarrow \infty), 
\label{dil-finalcorr}
\ee

\noi
where we have included the explicit asymptotic behavior for later 
usage.  The general solution of Eq. (\ref{dil-finalcorr}) is a particular 
solution of the inhomogeneous equation, plus the fundamental solution 
(\ref{fund-Lap}) with the arbitrary constants chosen to satisfy the 
boundary conditions. The general solution for $\psi_{1}(r)$,

\be
\psi_{1}(r) = \log(r) \int r \hspace{4pt} 
\textnormal{sech}^{2}(\frac{r}{\sqrt{2}}) dr - 
\int r \log(r) \textnormal{sech}^{2}(\frac{r}{\sqrt{2}}) dr + C_{1} + 
C_{2} \log(r),
\ee

\noi
with the requirement that $\psi_{1}(\infty)=0$.  Evaluating the first 
integral explicitly, and then integrating the second 
integral by parts using the result just obtained, allows us to bring 
this to the much more convenient form,

\be
\psi_{1}(r) = \left. \int_{a}^{r} \left[ \sqrt{2} \tanh(\frac{x}
{\sqrt{2}}) - 2 
\frac{\log[\cosh(\frac{x}{\sqrt{2}})]}{x} \right] dx \right. + C_{1} + 
C_{2} \log{r},
\label{dil-soltn1}
\ee

\noi
where we have introduced a lower integration limit $a$, to be 
determined momentarily.  In order to be able to impose the boundary 
condition 
$\psi_{1}(\infty)=0$, we need to understand the convergence of this 
integral as a (type I) improper integral.  It is easy to show that in 
fact the 
integral is logarithmically divergent as $r\rightarrow\infty$ since,

\be
\lim_{r\rightarrow\infty} \frac{\sqrt{2} \tanh(\frac{r}{\sqrt{2}}) - 2 
\frac{\log[\cosh(\frac{r}{\sqrt{2}})]}{r}}{\frac{1}{r}} = 2 \log(2).
\ee

\noi
If we rewrite the integrand in terms of exponentials, this limit is 
made
more evident, as well as allowing us to write a closed form 
expression 
for the integral.  Denoting the integrand by $F(r)$ we have

\bea
F(r) &=& \sqrt{2} \left[ \frac{1 - e^{-\sqrt{2} r}}{1+e^{-\sqrt{2}r} } 
\right]  
 - \sqrt{2} - \frac{2 \log(1+e^{-\sqrt{2}r})}{r} + \frac{2 
\log(2)}{r} \nonumber \\
&=& 2 \sqrt{2} \sum^{\infty}_{n=1} (-1)^{n} e^{-n\sqrt{2}r} - \frac{2}
{r} \sum^{\infty}_{n=1} (-1)^{n+1} \frac{e^{-n \sqrt{2} r}}{n} + 
\frac{2\log(2)}{r}, 
\label{integrand}
\eea

\noi
whence it is clear that the last term yields the logarithmic 
divergence, whereas the other terms yield obviously convergent 
integrals.  
This divergence must be canceled by the $C_{2} \log(r)$ term
of the homogeneous solution (\ref{fund-Lap}), by setting $C_{2} = -2 
\log(2)$.  This is a necessary condition of being able to 
impose $\psi_{1}(\infty)=0$.  Then, pulling the homogeneous solution 
$-2\log(2) \log(r)$ under the integral to cancel the $2\log(2)/r$ 
piece, to fully impose 
the boundary condition we must take the integration limit $a$ to 
infinity since 
the integrand is monotonic.  Also, we must take the constant 
homogeneous solution $C_{1}=0$.  Putting it all together, we finally 
have

\bea
\psi_{1}(r) &=& \left. \int^{r}_{\infty} \left[ 2 \sqrt{2} 
\sum^{\infty}_{n=1} 
(-1)^{n} e^{-n\sqrt{2}r} - \frac{2}{r} \sum^{\infty}_{n=1} 
(-1)^{n+1} \frac{e^{-n \sqrt{2} r}}{n} \right] dr \right. \nonumber \\
&=& 2 \log(1+ e^{-\sqrt{2}r}) + 2 \sum_{n=1}^{\infty} 
\frac{(-1)^{n+1}}{n} 
\textnormal{Ei}_{1}(n\sqrt{2}r),
\label{dil-soltn2}
\eea

\noi
where we have introduced the exponential integral defined by

\be
\textnormal{Ei}_{1}(x) = \int_{1}^{\infty} \frac{e^{-x t}}{t} dt.
\ee

\noi
It is easy to verify explicitly that this solves the dilaton 
correction equation (\ref{dil-finalcorr}) and satisfies

\be
\lim_{r\rightarrow\infty} \psi_{1}(r) = 0.
\ee

\noi 
However, though we have been able set the dilaton $\psi$ equal to 
zero at 
spatial infinity, the dilaton now diverges to $+\infty$ at $r=0$ since

\bea
\lim_{r\rightarrow 0} \psi_{1}(r) &=& \lim_{r\rightarrow 0} 
2 \sum_{n=1}^{\infty} \frac{(-1)^{n+1}}{n} 
\textnormal{Ei}_{1}(n\sqrt{2}r) \sim \lim_{r\rightarrow 0}
-2 \sum_{n=1}^{\infty} \frac{(-1)^{n+1}}{n}
\log(r) \nonumber \\
&=&  \lim_{r\rightarrow 0} - 2 \log(2) \log(r) \rightarrow +\infty,
\eea

\noi
using the fact that

\be
\lim_{r\rightarrow 0} \frac{\textnormal{Ei}_{1}(a r)}{- \log(r)} = 
\lim_{r\rightarrow 0} \frac{ \frac{-e^{-a r}}{r} }{ \frac{-1}{r} } = 1 
\hspace{10pt} \forall a>0.
\ee

\noi
How did this come about?  This singularity is none other than the 
one introduced when we were forced to assign a nonzero value to the 
homogeneous term $C_{2} \log(r)$ in order to obey the boundary 
condition at infinity.  Thus in order to avoid a logarithmic 
divergence at infinity, we are forced to introduce one at zero by 
turning on $\log(r)$.  This can be viewed as a direct consequence of 
the fact that we are dealing with an essentially two-dimensional 
problem and the two-dimensional Laplace equation.\\

\noi
It is now clear why this result is independent of the 
parametrizations (\ref{h-par}),(\ref{g-par}), and of the truncation 
made in going to Eq. (\ref{dil-finalcorr}).  The $C_{2} \log(r)$ 
homogeneous 
term is turned on (and effectively shifts the particular solution) if 
and only if 
the (unshifted) particular solution integral is asymptotically 
divergent, which in 
turn depends only on the dominant asymptotic behavior of the 
Nielsen-Olesen 
solutions.  But this is precisely how we chose the parametrization 
and made the truncation: they have the correct asymptotic behavior.  
Conversely, once the $C_{2} \log(r)$ term is turned on, we now 
unavoidably have a positive logarithmic divergence at $r=0$, because 
the {\it unshifted} integrand is well behaved near $r=0$.  Again, we 
chose our parametrization to have the correct small r behavior of the 
Nielsen-Olesen solutions.\\

\noi
Finally one might worry in taking, as most authors including 
Nielsen and Olesen do, the asymptotic behavior of $\phi_{0}$ as 
$\exp(-\sqrt{2} r)$ and not $\exp(-\sqrt{2} r)/\sqrt{r}$, 
that we may have affected the convergence of the unshifted particular 
integral.  This is not the case.  Proceeding exactly as above, and
retaining only the dominant asymptotic contribution, it is easy to 
show that the boundary condition at infinity again forces us to turn 
on the homogeneous solution.  We worked with a simpler global 
parametrization before, so that we could discuss small r behavior of 
the solution as well.\\

\section{Discussion}

\noi
The results of the previous section are perhaps surprising.  In fact, 
this is a rather generic property of solutions to the inhomogeneous 
equation

\be
\psi_{1} ^{\prime\prime} + \frac{\psi_{1}^{\prime}}{r} = f(r)
\label{gen-inhomo-eqn}
\ee

\noi
with a vanishing asymptotic boundary condition, and with reasonable 
assumptions on $f(r)$.  As we have seen, the general solution of 
Eq. (\ref{gen-inhomo-eqn}) can be written as

\bea
\psi_{1}(r) &=& \log(r) \int r f(r) dr - \int r \log(r) f(r) dr + 
C_{1} + 
C_{2} \log(r)  \nonumber \\
&=& \log(r) \int_{a}^{r} x f(x) dx - \int_{b}^{r} x \log(x) f(x) 
dx,    
\label{gen-soltn1}
\eea

\noi 
where we have absorbed the homogeneous solution into the particular 
indefinite integrals by making them definite integrals: the arbitrary 
constants of the general solution are now the lower, constant, limits 
of integration.  Clearly, we cannot in general impose the boundary 
condition $\psi_{1}(\infty)=0$.  A necessary condition for being able 
to impose this condition is that

\be
\lim_{r \rightarrow \infty} \int^{r} x \log(x) f(x) dx
\label{fund-int1}
\ee

\noi
exists.  Unfortunately, this is not quite sufficient ($f(x) = 
\sin(x^{2})/[x \log(x)]$ furnishes a counterexample). 
However, the {\it absolute} convergence of the integral 
(\ref{fund-int1}) 
is sufficient to be able to impose $\psi_{1}(\infty)=0$, i.e. if

\be
\lim_{r \rightarrow \infty} \int^{r} x \log(x) \left| f(x) \right| 
dx = K < \infty.
\label{fund-int2}
\ee

\noi
For if this limit exists, then so does the limit

\be
\lim_{r \rightarrow \infty} \int^{r} x \left| f(x) \right| dx.
\ee

\noi
Then the squeeze theorem and the inequalities

\be
0 \le \left| \log(r) \int_{r}^{\infty} x f(x) dx \right| 
\le \int_{r}^{\infty} \log(r) x \left| f(x) \right| dx 
\le \int_{r}^{\infty} \log(x) x \left| f(x) \right| dx
\rightarrow 0 \hspace{6pt} as \hspace{6pt} r \rightarrow \infty
\ee

\noi
imply that

\be
\lim_{r\rightarrow\infty} \log(r) \int_{r}^{\infty} x f(x) dx = 0.
\label{int-cond}
\ee

\noi
This establishes the sufficiency of the condition (\ref{fund-int2}).\\

\noi
From Eq. (\ref{dil-1storder}), the actual $f(r)$ in which we are 
interested is 
determined from the Nielsen-Olesen solutions $\phi_{0}$ and $v_{0}$, 
and the 
arguments from the previous section establish that this $f(r)$ 
decays exponentially as $r\rightarrow \infty$.  Thus we easily satisfy
the above sufficient condition allowing us to take 
$\psi_{1}(\infty)=0$.\\

\noi
Now consider the behavior of $\psi_{1}(r)$ near $r=0$, {\it 
subsequent} to imposing $\psi_{1}(\infty)=0$.  We now write the 
solution (\ref{gen-soltn1}) as

\be
\psi_{1}(r) = \int_{r}^{\infty} x \log(x) f(x) dx - 
\log(r) \int_{r}^{\infty} x f(x) dx.
\ee

\noi
Remembering that $x \log(x) \rightarrow 0$ as $x\rightarrow 0^{+}$, 
we 
now demonstrate the inevitable presence of a logarithmic divergence 
of $\psi_{1}(r)$ at $r=0$ as 
long as $f(r)$ is well behaved near $r=0$ and $K \equiv 
\int_{0}^{\infty} x 
f(x) dx \ne 0$.  The sign of the divergence will depend on the sign 
of K.  Explicitly we have

\be
\lim_{r\rightarrow 0} \psi_{1}(r) \sim  \int_{0}^{\infty} x \log(x) 
f(x) dx - 
\log(r) \int_{0}^{\infty} x f(x) dx  \rightarrow  \textnormal{sgn}(K) 
\cdot \infty.   
\ee

\noi
Note that these integrals exist assuming only, in addition to the 
previous restrictions on f ensuring improper convergence, that f is 
defined 
and say continuous (or Riemann integrable) everywhere on $r\ge 0$, 
and in particular at 0. 
\footnote{Of course if f is poorly behaved (say divergent) as 
$r\rightarrow 0$, 
so that the integral diverges, then already the dilaton diverges 
without further argument.}\\

\noi
Again, because our $f(r)$ from Eq. (\ref{dil-1storder}) is defined and 
continuous 
for all $r\ge 0$ because the Nielsen-Olesen solutions are [remember 
that the term 
$(v_{0}^{\prime})^{2}/r^{2}$ in Eq. (\ref{dil-1storder}) is finite as 
$r\rightarrow 
0$ as seen in Eq. (\ref{dil-nexttofinalcorr}); in other words the field 
strength of 
the Nielsen-Olesen vortex is finite at the core], we have a 
logarithmic divergence at $r=0$ as explicitly shown in the previous 
section.  
In fact, since our $f(r)$ is explicitly non-negative (as seen in either 
Eq. (\ref{dil-nexttofinalcorr}) 
or its truncation (\ref{dil-finalcorr})), the K defined above is 
positive, and so the logarithmic divergence is to {\it positive} 
infinity at $r=0$.  Again, this was seen explicitly in the last 
section.\\

\noi
To summarize, we have found that a solution to Eq. (\ref{gen-inhomo-eqn}) 
can satisfy $\psi_{1}(\infty)=0$, if the limit (\ref{fund-int2}) 
exists.
Furthermore, if this limit exists so that we may impose
$\psi_{1} (\infty)=0$, the solution diverges logarthmically at 
$r=0$.  
Thus $\psi(\infty)=0$ implies $\psi(0)=\infty$.  Since the $f(r)$ 
relevant to our discussion decays exponentially as 
$r\rightarrow\infty$, 
and is well behaved at $r=0$, this is provides a general and generic 
proof of our result.
Incidentally, this also shows why our results of the previous section 
are independent of 
either the parametrizations to the Nielsen-Olesen solutions or the 
truncation made in going from Eq. (\ref{dil-nexttofinalcorr}) to 
Eq. (\ref{dil-finalcorr}): this general behavior depends only on the 
behavior of f as $r\rightarrow \infty$ and as $r\rightarrow 0$, and 
our 
parametrization was chosen to be exact in these limits.\\

\noi
Given that we have now established that this dilaton behavior is 
rather 
generic, one might wonder if this divergent behavior of the dilaton 
at the core of the vortex is somehow an artifact of the perturbation 
theory.  
In fact, we now expect the full dilaton equation to yield even worse 
behavior 
because of the exponential feedback.  As a consistency check of our 
result, we will briefly examine the full dilaton equation 
(\ref{d-ode2}).  
If we take the perturbation theory to be valid only for very large r, 
where the 
dilaton VEV is still small, so that we are still in a classical 
and perturbative regime, we know that it starts to run positive as 
one 
comes in from spatial infinity.  A positive exponential self-coupling 
acts 
as a source term that becomes larger and larger as $r\rightarrow 0$.  
So if we equate small $r$ with large $\psi$, then the dilaton 
equation (\ref{d-ode2}) is dominated by the vacuum Fayet-Iliopoulos 
term \cite{p} proportional to $e^{6\psi}$ [or $1/(S+S^{\dagger})^{3}$ 
in the 
notation of Polchinski], which comes directly from the anomaly 
cancellation as a 
two string-loop tadpole \cite{dsw}, so that, approximately

\be
\psi^{\prime\prime} + \frac{\psi^{\prime}}{r} \sim \frac{3\beta}{2} 
e^{6\psi},
\label{large-d-approx}
\ee

\noi
where we are taking $\beta$ so small that we can neglect the axion 
contribution that is otherwise possibly as large (but of the same 
sign 
in any case), and where we are assuming that we still have $\phi 
\rightarrow 0$ as $r\rightarrow 0$;
i.e. the vortex is well defined.  An exact solution to 
Eq. (\ref{large-d-approx}) is given by 

\be
\psi(r) \sim \frac{-1}{6} \log\left[a_{1}r \left(1- \frac{9\beta}
{2 a_{1} } 
r\right)^{2} \right],    
\ee

\noi
where $a_{1}$ is an undetermined constant.  For very small $\beta$ 
this is essentially the same behavior as our perturbative calculations.  
This solution is obviously consistent with the approximation 
(\ref{large-d-approx}) to the full dilaton equation (\ref{d-ode2}) 
if we assume that the gauge field and Higgs boson still have the 
boundary values $\phi(0)=0$, and $v^{\prime}(0)=0$.\\

\noi
In any case, we seem to be led to the conclusion that the 
4-dimensional dilaton in this model starts to grow as we come in from 
spatial infinity.  Since the dilaton VEV in this model sets the 
anomalous U(1) gauge coupling,  we eventually enter a strongly 
coupled regime where not only the $\beta$ perturbation theory breaks down, 
but where it no longer makes sense to ignore quantum and string 
threshold corrections.  In other words, such a vortex is 
fundamentally 
a quantum mechanical object.  Furthermore, as we have seen, the 
unavoidable 
singularities we have encountered are a direct consequence 
of the effectively {\it two}-dimensional nature of the vortex system: 
the solution of the Laplace (or Poisson) equation in two dimensions 
involves a logarithm which is singular at both $r=0$ and 
$r\rightarrow\infty$.\\  

\noi
Our conclusion then is that anomalous U(1) vortex solutions of 
heterotic superstring 
theory, if they are to have the standard asymptotic structure at 
large radial distances 
from the vortex core, necessarily generate large dilaton field values 
within that core 
signaling the presence of strong coupling and large quantum 
fluctuations.  As such, 
these vortices can never be adequately described as entirely 
classical objects; their 
classical exterior surrounds an interior that is intrinsically 
quantum mechanical.


\vspace{20pt}
\noindent{ {\bf Acknowledgments} } \\
\noindent This work was supported in part by the Natural Sciences and 
Engineering Research  
Council of Canada.

\end{document}